\documentclass[a4paper,10pt,twoside]{article}

\usepackage[english]{babel}
\usepackage{graphicx}
\usepackage{sidecap}
\usepackage{fancyhdr}
\usepackage{ifthen}
\usepackage[official,right]{eurosym}
\usepackage{wrapfig,subfigure}
\usepackage[usenames,dvipsnames]{color}
\usepackage{tabularx}
\usepackage{amsmath,amssymb}
\usepackage{caption}
\usepackage{array,dcolumn}
\usepackage{helvet}
\usepackage[colorlinks=true,allcolors=blue]{hyperref}
\usepackage{lastpage}

\DeclareMathAlphabet{\mathbfit}{OT1}{cmr}{bx}{it}

\begin{document}

\fancyhf{}
\fancyhead[LE]{\nouppercase{\it \leftmark}}
\fancyhead[RO]{\nouppercase{\it \leftmark}}
\fancyfoot[C]{Page \thepage\ of \pageref{LastPage}}

\thispagestyle{empty}
\begin{tabular}{l}
\Large Final Report for a\\
\end{tabular}
\begin{figure}[!h]
\begin{tabular}{l}
\resizebox{0.55\columnwidth}{!}{
\includegraphics[width=0.6\textwidth]{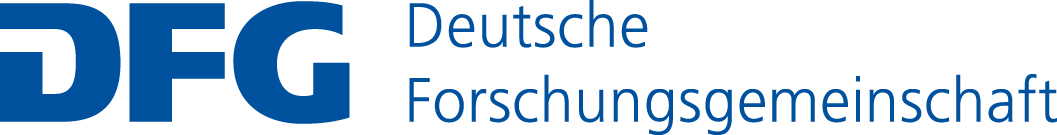}}
\vspace{-10pt}
\end{tabular}
\end{figure}

\begin{tabular}{l}
\Large Research Grant ~~~~~~~~~~~~~~~~~~~~~~~~~~~~~~~~~~~~~~~~~~~~~~~~~~~~~~~~~~~~~~~~~~~~~~~~~~~~~~~~~~~~~\\
\\
\hline
\end{tabular}

\vspace{4 cm}

\begin{center}
\huge\bf Druckabh\"{a}ngige Untersuchung eisenbasierter Supraleiter mittels Kernspinresonanz \\

\vspace{0.5cm}
\Large Applicant: Dr. {Adam P. Dioguardi}
\end{center}

\vspace{9.5cm}

\begin{tabular}{l}
\hline
\\
\Large Leibniz-Institut f\"{u}r Festk\"{o}rper- und Werkstoffforschung~~~~~~~~~~~~~~~~~~~~~~~~\\
\\
\Large  IFW Dresden~~~~~~~~~~~~~~~~~~~\\
\\
\end{tabular}

\newpage


\section{General Information}
\label{general_information}

\begin{enumerate}
\item DFG reference number: DI2538/1-1
\item Project number: 418764509
\item Project title: Druckabh\"{a}ngige Untersuchung eisenbasierter Supraleiter mittels Kernspinresonanz
\item Name(s) of the applicant(s): Adam P. Dioguardi
\item Official address(es):
    
    Leibniz-Institut f\"{u}r Festk\"{o}rper- und Werkstoffforschung Dresend (IFW) e.V.\\
    Helmholtzstra{\ss}e 20\\
    01069 Dresden
\item List of the maximum ten most important publications to emerge from this project:

    \begin{enumerate}
        
        \item 
            I. Jakovac, \textbf{A. P. Dioguardi}, M. S. Grbi\'{c}, G. D. Gu, J. M. Tranquada, C. W. Hicks, M. Po\v{z}ek, and H.-J. Grafe, ``Uniaxial stress study of spin and charge stripes in La$_{1.875}$Ba$_{0.125}$CuO$_4$ by ${}^{139}$La NMR and ${}^{63}$Cu NQR,'' \href{https://doi.org/10.1103/PhysRevB.108.205113}{Physical Review B \textbf{108}, 205113 (2023)}.
        \item 
            F. Bougamha, S. Selter, Y. Shemerliuk, S. Aswartham, A. Benali, B. B\"{u}chner, H.-J. Grafe, and \textbf{A. P. Dioguardi}, ``${}^{31}$P NMR investigation of quasi-two-dimensional magnetic correlations in $T_2$P$_2$S$_6$ ($T$ = Mn, Ni),'' \href{https://doi.org/10.1103/PhysRevB.105.024410}{Physical Review B \textbf{105}, 024410 (2022)}.
        \item 
            P. Lepucki, R. Havemann, \textbf{A. P. Dioguardi}, F. Scaravaggi, A. U. B. Wolter, R. Kappenberger, S. Aswartham, S. Wurmehl, B. B\"{u}chner, and H.-J. Grafe, ``Mapping out the spin fluctuations in Co-doped LaFeAsO single crystals by NMR,'' \href{https://doi.org/10.1103/PhysRevB.103.L180506}{Physical Review B \textbf{103}, L180506 (2021)}.
        \item 
            \textbf{A. P. Dioguardi}, S. Selter, U. Peeck, S. Aswartham, M.-I. Sturza, R. Murugesan, M. S. Eldeeb, L. Hozoi, B. B\"{u}chner, and H.-J. Grafe, ``Quasi-two-dimensional magnetic correlations in Ni$_2$P$_2$S$_6$ probed by 31P NMR,'' \href{https://doi.org/10.1103/PhysRevB.102.064429}{Physical Review B \textbf{102}, 064429 (2020)}.
    \end{enumerate}
\end{enumerate}

\section{Progress Report}
\label{progress_report}

\subsection{Overall progression of the work}
\label{progression_of_work}

Although the project goals had to be significantly modified as challenges arose and reassesments had to be made, this DFG Eigenestelle grant supported the publication of four impactful peer-reviewed publications, development and improvement of the strain-tuned NMR capability at IFW Dresden, supported the development of a number of simulation and automation software packages, and supported the PI during mentorship of several PhD, masters, and bachelors students.

The project immediately hit some technical challenges that had to be overcome before we could successfully conduct high signal-to-noise ratio NMR experiments on samples small enough to approach the 1\% strain requirements to induce an ostensible spin reorientation of the Fe spins in the spin-density wave state of BaFe$_2$As$_2$. The biggest challenge was getting high enough signal-to-noise ratio for appropriately sized samples. Although the existing strain cells had already been used, the required signal-to-noise ratio for the planned experiments, especially in the magnetic state, meant that each spectrum required several days to collect. Each time a sample fractured, we had to start over from scratch, resulting in significant setbacks.

Once enough data was collected we found that our measurements were inconsistent with previous measurements from the second-generation strain cell. After many attempts to reproduce the original measurements and those from the literature, the source of this issue was found to be significant differences in the assumed calibration of the cells. Therefore we used a precision optical interferometer to calibrate of our second-generation cells. Some of the differences also came from deformation of the mounting epoxy, and a scheme was developed based on the literature to account for this reduced displacement in comparison to the value extracted from the capacitive dilatometer. Simultaneously, we utilized funds from this grant to develop software to control the new control hardware, and finally to automate measurements and optimize experiment times.

Once these issues were sufficiently accounted for, systematic ${}^{75}$As nuclear magnetic resonance (NMR) experiments were conducted in BaFe$_2$As$_2$ under in-situ-controlled uniaxial pressure. The electric field gradient (EFG), spin-lattice relaxation rate ($T_1^{-1}$), spin-spin relaxation rate ($T_2^{-1}$), and the Knight shift $K$ at the As site are sensitive to applied uniaxial pressure (the second two points of which are novel, and the first agrees with previous published results). Uniaxial pressure increases, broadens, and separates the N\'{e}el temperature $T_N$ and the tetragonal-to-orthorhombic structural transition $T_S$. However, spectral measurements in the magnetic state find no evidence for spin polarization reorientation for applied uniaxial pressure up to the point of sample failure; the maximum value achieved was $\varepsilon = -0.009$. A manuscript for this work was written, but due to the main result of lack of spin reorientation, the manuscript has not yet been published, because we decided to focus on other materials on which we hoped publication would be more impactful.

Unfortunately, we were unable to perform successful strain-dependent NMR measurments of CaFe$_2$As$_2$, because the samples were too ductile and/or micaceous---i.e. the layers of the crystals were not as strongly bound together along the c-axis direction as in BaFe$_2$As$_2$---and all samples that we attempted to measure broke before we could complete systematic measurements.

The Cs- and Rb-based 122 materials and 1111 materials were not investigated due to our collaborators' difficulty with synthesis of large-enough samples for the strain measurements~\cite{Kappenberger_2018_Solidstatesingle}, though we were able to contribute to the understand of the F- and Co-doped 1111 Fe-based materials and compare them to the 122s~\cite{Lepucki_2021_Mappingoutspin}, though not under conditions of uniaxial pressure.

Instead, we focused on LiFeAs, which was successfully mounted in the strain cell in a glove box and coated with a thin layer of epoxy to protect the aire-sensitive samples during transfer to the NMR cryostat's inert-gas environment. We were able to successfully measure the strain dependence of LiFeAs, but the nematic susceptibility, as evidenced by strain dependence of the electric field gradient, were so small in comparison to BaFe$_2$As$_2$, that we decided that these results were unable to probe enough interesting physics. Therefore, we modified our plans and focused on magnetism in the 226 materials, also thought to be candidates in which strain tuning of the ground state is possible \cite{Grasso_1986_Opticalabsorptionspectra, Wildes_2015_Magneticstructurequasi, Kim_2019_MottMetalInsulator}. 

We performed initial measurements on two such compounds, which turned out to be interesting enough to investigate first in depth without strain tuning. First, detailed ${}^{31}$P nuclear magnetic resonance (NMR) measurements are presented on well characterized single-crystals of antiferromagnetic van der Waals Ni$_2$P$_2$S$_6$. We observed an anomalous breakdown in the proportionality of the NMR shift $K$ with the bulk susceptibility $\chi$. This so-called $K$--$\chi$ anomaly occurs in close proximity to the broad peak in $\chi(T)$, thereby implying a connection to quasi-2D magnetic correlations known to be responsible for this maximum. Quantum chemistry calculations show that crystal field energy level depopulation effects cannot be responsible for the $K$--$\chi$ anomaly. Appreciable in-plane transferred hyperfine coupling is observed, which is consistent with the proposed Ni--S--Ni super- and Ni--S--S--Ni super-super-exchange coupling mechanisms. Magnetization and spin-lattice relaxation rate ($T_1^{-1}$) measurements indicate little-to-no magnetic field dependence of the N{\'e}el temperature. Finally, $T_1^{-1}(T)$ evidences relaxation driven by three-magnon scattering in the antiferromagnetic state~\cite{Dioguardi_2020_Quasitwodimensional}.

We then reported the anomalous breakdown in the scaling of the microscopic magnetic susceptibility---as measured via the ${}^{31}$P nuclear magnetic resonance (NMR) shift $K$---with the bulk magnetic susceptibility $\chi$ in the paramagnetic state of Mn$_2$P$_2$S$_6$. This anomaly occurs near $T_\mathrm{max} \sim 117$\,K the maximum in $\chi(T)$ and is therefore associated with the onset of quasi-two-dimensional (quasi-2D) magnetic correlations. The spin--lattice relaxation rate divided by temperature $(T_1T)^{-1}$ in Mn$_2$P$_2$S$_6$ exhibits broad peak-like behavior as a function of temperature, qualitatively following $\chi$, but displaying no evidence of critical slowing down above the N\'{e}el temperature $T_N$.  In the magnetic state of Mn$_2$P$_2$S$_6$, NMR spectra provide good evidence for 60 degree rotation of stacking-fault-induced magnetic domains, as well as observation of the spin-flop transition that onsets at 4\,T. The temperature-dependent critical behavior of the internal hyperfine field at the P site in Mn$_2$P$_2$S$_6$ is consistent with previous measurements and the two-dimensional anisotropic Heisenberg model. In a sample of Ni$_2$P$_2$S$_6$, we observe only two magnetically split resonances in the magnetic state, demonstrating that the multiple-peaked NMR spectra previously associated with 60 degree rotation of stacking faults is sample dependent. Finally, we report the observation of a spin-flop-induced splitting of the NMR spectra in Ni$_2$P$_2$S$_6$, with an onset spin-flop field of $\mu_0 H_\mathrm{sf} = 14$\,T~\cite{Bougamha_2022_31PNMRinvestigation}.

And finally, when we could resume strain measurements, we focused on LBCO strain-tuned NMR, where we studied the response of spin and charge order in single crystals of La$_{1.875}$Ba$_{0.125}$CuO$_4$ to uniaxial stress, via $^{139}$La nuclear magnetic resonance (NMR) and $^{63}$Cu nuclear quadrupole resonance (NQR), respectively. In unstressed La$_{1.875}$Ba$_{0.125}$CuO$_4$, the low-temperature tetragonal structure onsets below $T_\text{LTT} = 57$\,K, while the charge order and the spin order transition temperatures are $T_\text{CO} = 54$\,K and $T_\text{SO} = 37$\,K, respectively. We find that uniaxial stress along the [110] lattice direction strongly suppresses $T_\text{CO}$ and $T_\text{SO}$, but has little effect on $T_\text{LTT}$. In other words, under stress along [110] a large splitting ($\approx$ 21 K) opens between $T_\text{CO}$ and $T_\text{LTT}$, showing that these transitions are not strongly associated with each other. Meanwhile, stress along [100] causes a slight suppression of $T_\text{LTT}$ but has essentially no effect on $T_\text{CO}$ and $T_\text{SO}$. Magnetic field H along [110] stabilizes the spin order: the suppression of $T_\text{SO}$ under stress along [110] is slower for H $\parallel$ [110] than H $\parallel$ [001]. We developed a Landau free energy model and interpret our findings as an interplay of symmetry breaking terms driven by the orientation of spins~\cite{Jakovac_2023_Uniaxialstressstudy}.

In this report we will focus on the measurements, analysis, and conclusions relevant to the proposed project. We hope to publish a manuscript on the results on BaFe$_2$As$_2$ in the near future that had been sidelined by the above-mentioned more successful studies.

\section{Experimental Infrastructure Advancements}
\label{sec:experimental_advancements}

To perform the measurements discussed above, we first needed to characterize our apparatus, solve issues with radio-frequency noise injected via the control leads, rebuild the NMR probe to allow for mounting of the two uniaxial pressure cells, to write software to control the electronics and automate measurements in the two compatible NMR systems. 

Software was written to wrap control of the following devices:
\begin{itemize}
    \item Andeen Haagerling capacitance bridge
    \item Keithley 2614B dual-channel high-voltage sourcemeter
    \item Arduino Due (driving Trek 610E voltage amplifier)
    \item Lake Shore 331 temperature controller
\end{itemize}

A graphical user interface, named pyStrainIFW, was then written to utilize the above wrappers to centralize control of all relevant parameters, protect the cells while loading and changing temperature, and logging all data other than the experimental NMR data.

Later, a package called AutoStrain was developed in collaboration with our Zagreb colleges, to automate measurements (when feasible/necessary) as a function of field, frequency, temperature, and applied uniaxial pressure.

Throughout the project, we also developed standard operating procedures to safely use the delicate uniaxial pressure cells. For example, the procedure for preparing and mounting samples that was developed throughout the project, via trial and error, is still in use today. It was also determined that the uniaxial pressure cells needed to be logged and monitored at all times, because loss of temperature control, static discharge, and perpendicular force applied to the piezo stacks, were all found to be drivers of cell failure.

\begin{figure}[h!] 
    \includegraphics[trim=0cm 0cm 0cm 0cm, clip=true, width=0.5\linewidth]{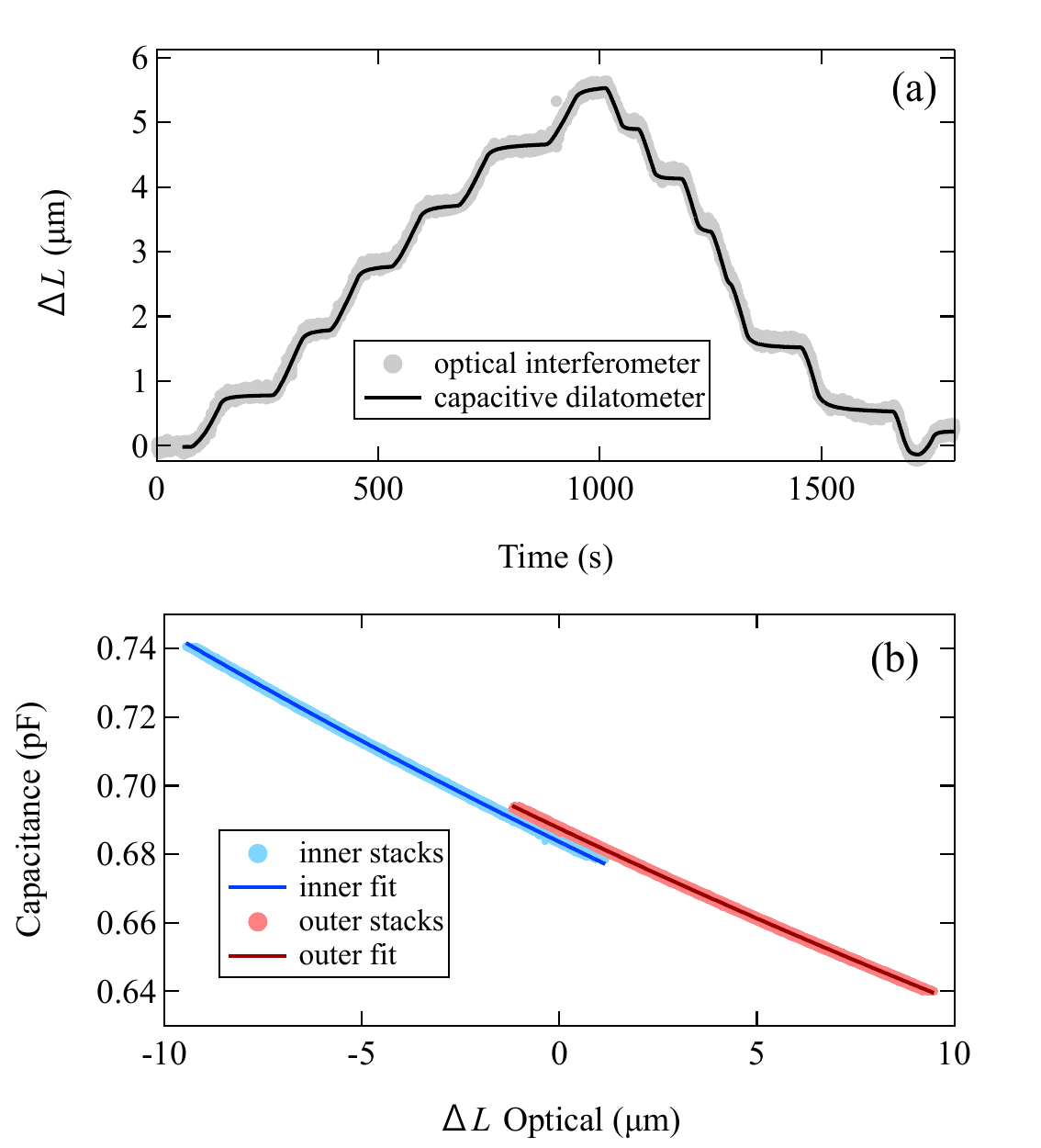}
    \includegraphics[trim=0cm 0cm 0cm 0cm, clip=true, width=0.5\linewidth]{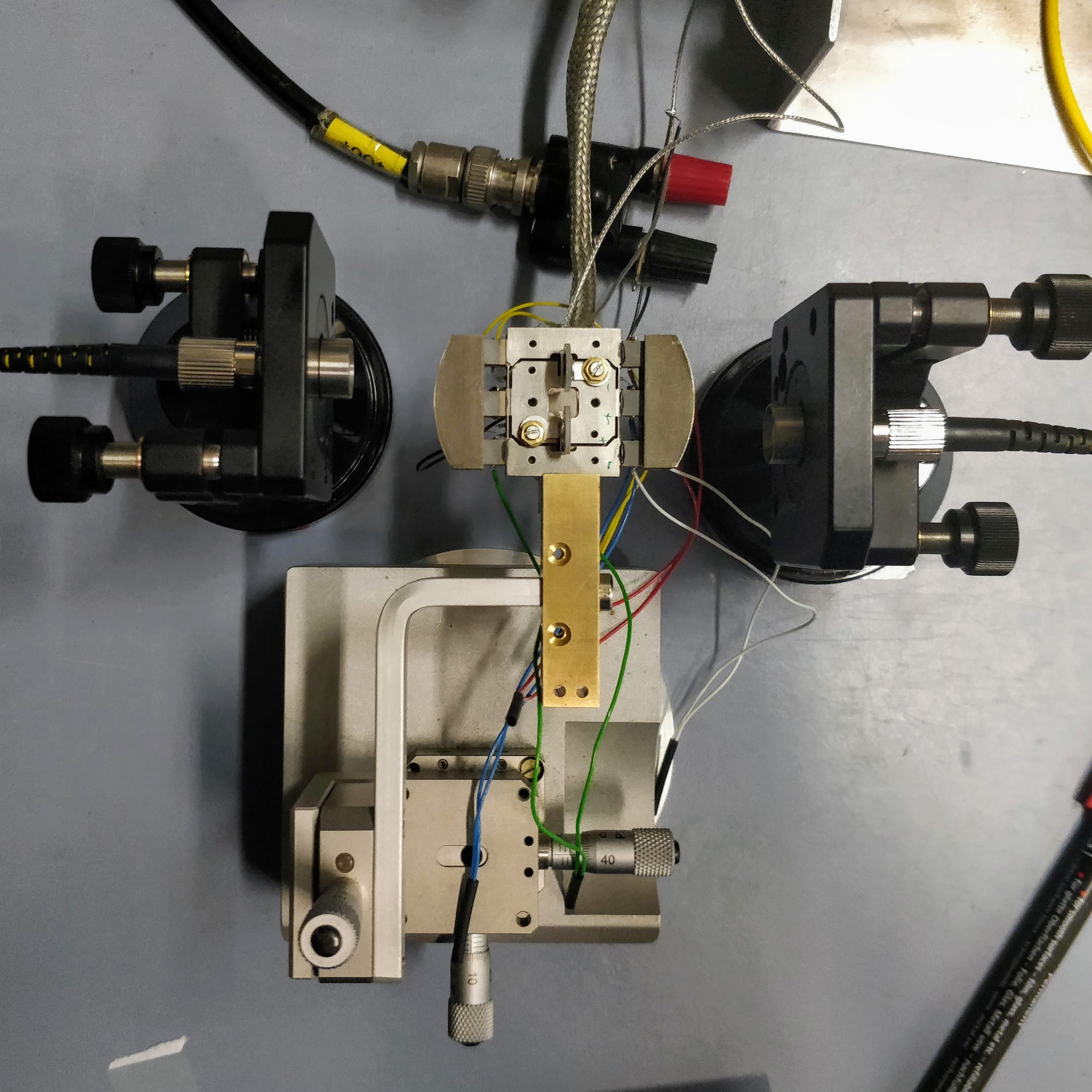}
    \caption{\label{fig:strain_cell_2_optical_calib_setup} The strain cell's capacitive dilatometer was calibrated by using a direct measurement of the change in distance as a function of applied voltage via the optical interferometer setup in the photo to the right. (a) Agreement between the measured change block position $\Delta L$ as measured by the optical interferometer and calibrated capacitive dilatometer. (b) Calibration curves of capacitance vs change in position as measured by the optical interferometer.}
\end{figure}

Optical interferometry was used to characterize the strain cells by measuring the relative displacement of the sample mounting blocks and the capacitance simultaneously. Figure~\ref{fig:strain_cell_2_optical_calib_setup}(a) shows the result of these measurements, i.e. we can accurately measure the relative displacement via the capacitive dilatometer as confirmed by the optical interferometer. We also learned that from run-to-run the offset of the curves is slightly different, as shown in Figure~\ref{fig:strain_cell_2_optical_calib_setup}(b). Our measurements do not rely on the ostensible zero point, but instead on intrinsic NMR parameters such as the asymmetry parameter of the electric field gradient tensor. 

The setup to measure the position of the two sample-mounting blocks via two lasers/spherical mirrors with optical interferometry is shown in Figure~\ref{fig:strain_cell_2_optical_calib_setup}. The relative motion is also measured by the capacitive dilatometer and we find that the known area of the capacitor plates $A=8.36$\,mm${}^2$ is within the uncertainty of the experiment-derived area $A_\text{exp} = 8.3 \pm 0.1$\,mm${}^2$. We therefore use the known area in our software to calculate the measured raw displacement, and then calculate the resulting corrected strain within the sample therefrom.




\section{Strain Correction Calculation}
\label{sec:strain_correction_calc}

In previous measurements, both at the IFW and as discussed with collaborators, the reported strain values were not properly corrected for loss. That is, some previous publications had likely over estimated the amount of strain in the sample for a measured displacement of the capacitive dilatometer.

Calculation of the corrected resultant strain in a sample is based on Hicks et al.~\cite{Hicks_2014_Piezoelectricbasedapparatus}. In this RSI the authors perform finite element analysis calculations to estimate the amount of transferred strain through the mounting epoxy. The relevant details are summarized in the following equations. First, the authors make an approximation for the change in force within the sample as a function of position:
\begin{equation}
    \label{eqn:force_in_sample}
    \frac{dF}{dx} \approx n w C_{66,e}\frac{D(x)}{d}, 
\end{equation}
where $n=2$ if the sample is bonded from both sides (or 1 if the sample is only bonded from one side), $w$ is the width of the sample, $C_{66,e}$ is the shear modulus of the epoxy, $D(x)$ is the displacement of the sample at position $x$ with respect to its unloaded position, and $d$ is the epoxy thickness. The solution to the differential equation for $\epsilon_{xx} = dD/dx$ can be solved via a decaying exponential characterized by the length scale over which the strain is transferred through the epoxy. This length scale is given by
\begin{equation}
    \label{eqn:strain_lambda}
    \lambda = \sqrt{\frac{Y t d}{n C_{66, e}}},
\end{equation}
where $Y$ is Young's modulus of the sample, $t$ is the sample thickness, $d$ is the epoxy thickness as mentioned above, $n=2$, and $C_{66,e}$ is the shear modulus of the epoxy. Using $\lambda$, we can calculate the estimated strain in the sample---based on the known capacitance dialtometry measurements---as the epoxy deforms via,
\begin{equation}
    \label{eqn:applied_corrected_strain}
    \epsilon = \frac{d - d_0}{L + 2\lambda},
\end{equation}
where $d$ is the measured displacement, $d_0$ is the initial displacement at zero strain, $L$ is the sample length between the plates, and $\lambda$ is the length scale over which the strain is transferred to the sample as given by Eqn.~\ref{eqn:strain_lambda}.

The most elusive part of the calculation input parameters was the determination of the Young's modulus $Y$ of the sample itself along the $\left[110\right]$ direction, given by:
\begin{equation}
Y_{\left[ 110 \right]} = 4\left( \frac{1}{C_{66}} + \frac{1}{ \frac{C_{11}}{2} + \frac{C_{12}}{2} - \frac{{C_{13}}^2}{C_{33}} } \right)^{-1},
\end{equation}
where $C_{ij}$ are the elements of the elastic tensor. We have derived this expression from the general case of the Young's modulus in arbitrary direction from Cazzani and Rovati, given by:
\begin{equation}
\frac{1}{Y(\mathbf{n})} = S_{11} - \left( S_{11} - S_{33} \right){n_3}^4 - \left( 2S_{11} - 2S_{12} - S_{66} \right){n_1}^2 {n_2}^2 - \left( 2S_{11} - 2S_{13} - S_{44} \right)\left( {n_1}^2 {n_3}^2 + {n_2}^2 {n_3}^2 \right),
\end{equation}
where $n_i$ are the components of unit vector $\mathbf{n}$ and $S_{ij}$ are the elements of the compliance tensor $\tilde{\mathbf{S}} = \tilde{\mathbf{C}}^{-1}$~\cite{Cazzani_2005_ExtremaYoungsmodulus}.

\begin{figure}[h!] 
\centering
    \includegraphics[trim=0cm 0cm 0cm 0cm, clip=true, width=0.9\linewidth]{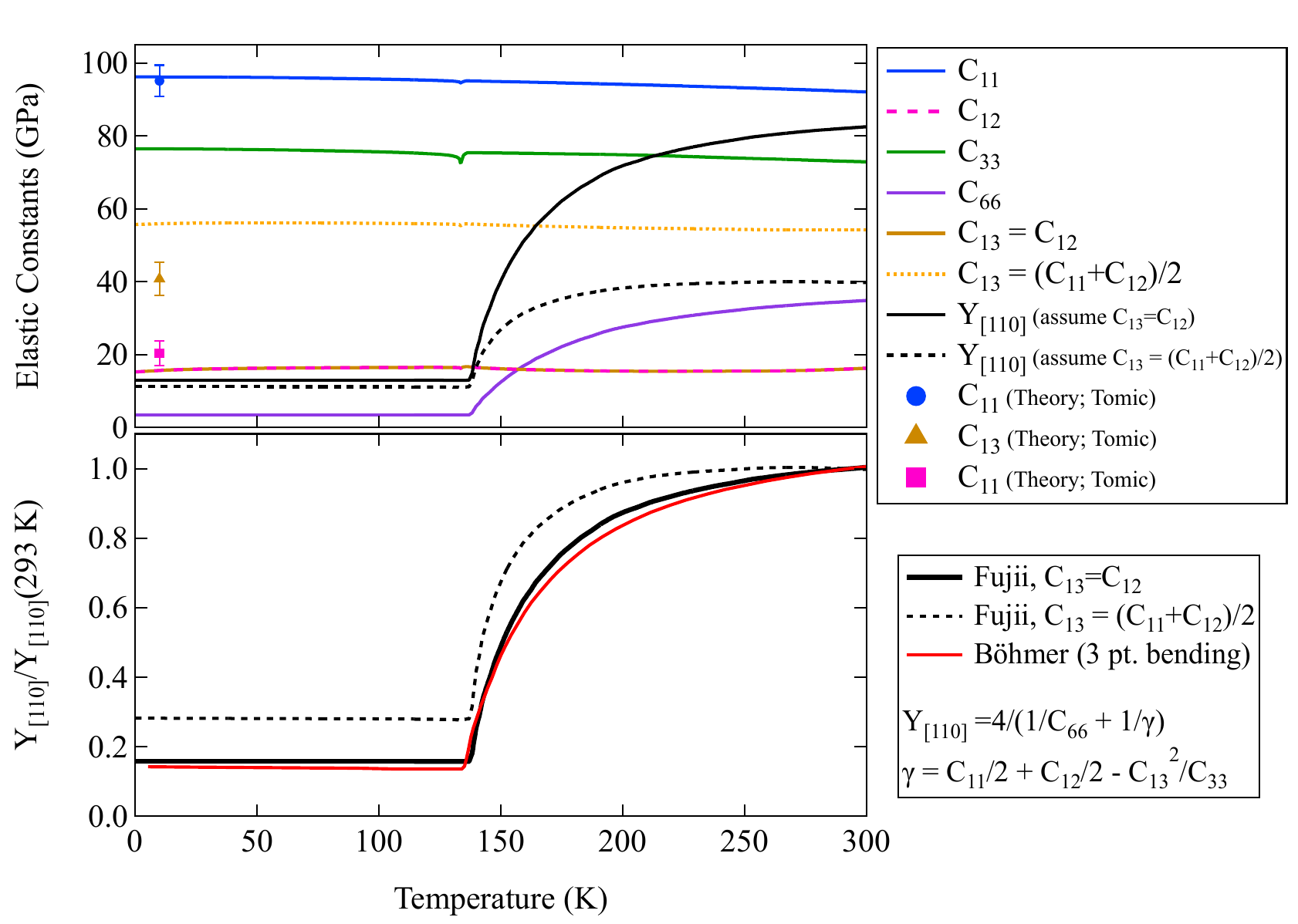}
    \caption{\label{fig_s1_Y110_normalization_compare} The top plot shows the temperature dependence of the elastic constants of BaFe$_2$As$_2$ as measured by Fujii~\cite{Fujii_2018_AnisotropicGrueneisenParameter}, with two reasonable assumed values for $C_{13}$ and the resulting Young's modulus along the [110] tetragonal direction. Markers indicate theoretical calculations of a subset of elastic constants from Tomi\'{c}~\cite{Tomic_2013_planeuniaxialstress}. The bottom plot shows the normalized Youngs modulus in comparison to that measured by B\"{o}hmer~\cite{Boehmer_2014_NematicSusceptibilityHole}.}
\end{figure}

B\"{o}hmer showed that $Y_{\left[ 110 \right]}$ has significant temperature dependence, but the quantity is normalized (see red curve in Fig.~\ref{fig_s1_Y110_normalization_compare}), and therefore not directly helpful for our strain-loss calculation~\cite{Boehmer_2014_NematicSusceptibilityHole}. Various elastic constants were measured by Fujii et al.~\cite{Fujii_2018_AnisotropicGrueneisenParameter}, except for the crucial elastic constant $C_{13}$, due to the experimental setup geometry. As a result we were forced to make an assumption, that either $C_{13} = C_{12}$, as Fujii et al. do, or

 
 
The elastic constants of epoxies have been relatively well studied as a function of temperature, with some small variability among different formulations~\cite{Hartwig_1978_LowTemperatureProperties}. For the above calculation we need to determine the shear modulus of the epoxy. The shear modulus of the epoxy $C_{66,e} = Y_e/(2(1+\nu_e))$, for an isotropic material, where $Y_e$ is Young's modulus along the direction of strain and $\nu_e$ Poisson's ratio~\cite{Landau_2012_TheoryElasticityVolume}. Hartwig did not study stycast specifically, but comments that all measured crosslinked epoxies had low temperature Poisson's ratios between 0.35 to 0.37~\cite{Hartwig_1978_LowTemperatureProperties}, therefore we will choose 0.36 as the Poisson ratio for the epoxy. They also comment that the Young's moduli are all within 10\% of each other, indicating that the low temperature values are roughly independent of chemical structure. We will use this comment as well as Fig. 7 from the same publication to argue that we can assume the temperature dependence of UHU endfest should scale with that of stycast-1266. UHU Endfest has $Y_E^\text{UHU} = 5.54$\,GPa and $\nu=0.35$ at room temperature~\cite{Voerden_2012_Springconstanttuning}. We also note that the Shore D hardness UHU Endfest is very close to that of Stycast-1266, 70 and 75, respecectively~\cite{Voerden_2012_Springconstanttuning, Henkel_2016_LoctiteStycast1266}.

\begin{figure}[h!] 
\centering
    \includegraphics[trim=0cm 0cm 0cm 0cm, clip=true, width=0.5\linewidth]{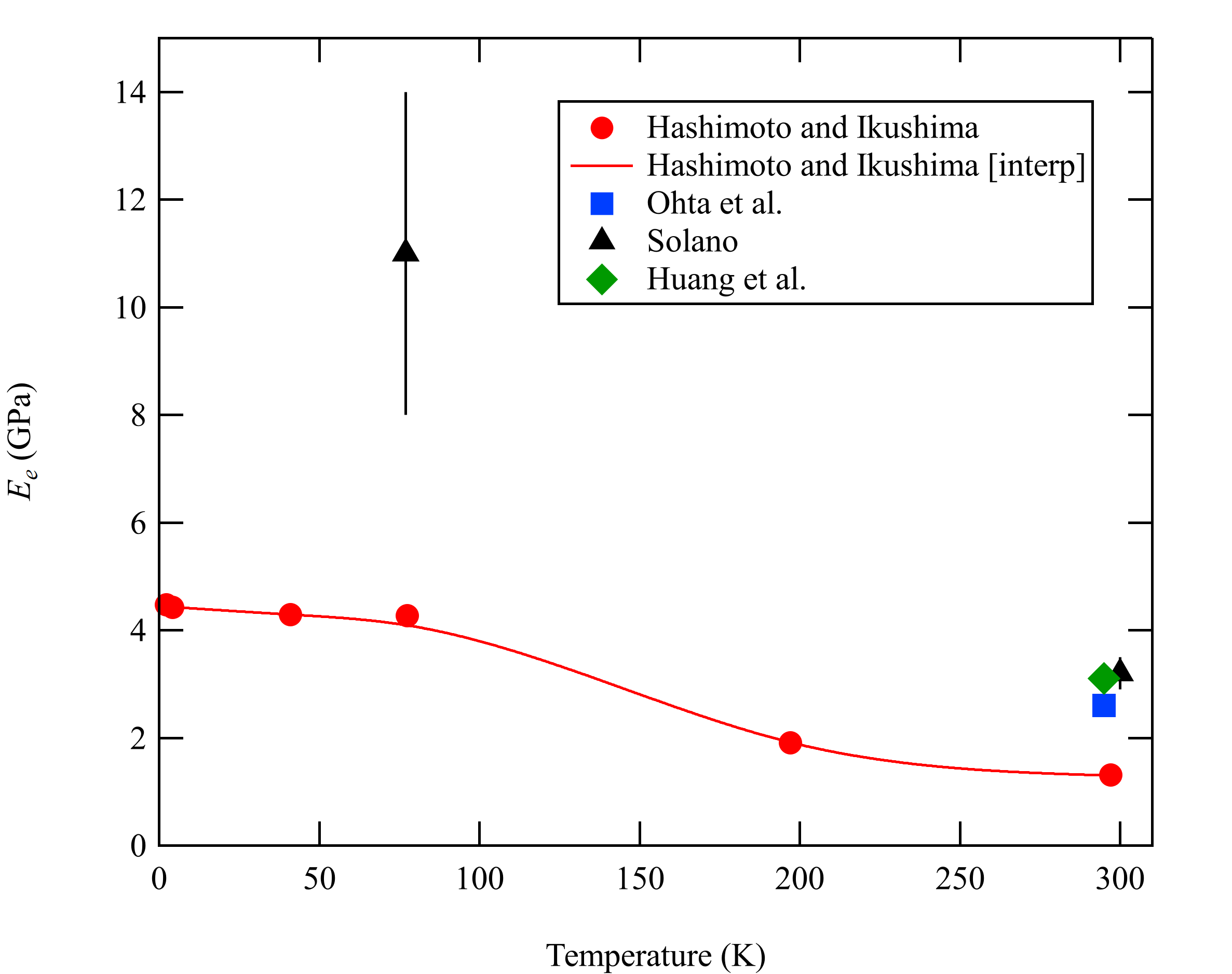}
    \caption{\label{fig:Ye_vs_temp_stycast1266} Temperature dependence of the Young's modulus of Stycast-1266 epoxy as a function of temperature~\cite{Hashimoto_1980_MechanicalpropertiesStycast, Ohta_2003_OriginPerpendicularMagnetic, Solano_2004_CryogenicSiliconMicrostrip, Huang_2002_MechanicalpropertiesZylon}.}
\end{figure}

Several studies exist in the literature~\cite{Hashimoto_1980_MechanicalpropertiesStycast, Solano_2004_CryogenicSiliconMicrostrip, Ohta_2003_OriginPerpendicularMagnetic, Huang_2002_MechanicalpropertiesZylon}, in which the Young's modulus of Stycast 1266 epoxy was measured at a variety of temperatures. The data are summarized in Figure~\ref{fig:Ye_vs_temp_stycast1266}.

In our measurements, the geometrical parameters of the samples were measured via calibrated microscope images. The the epoxy thickness was chosen by modifiing the thickness of the spacer plates. $d$ was measured using the calibrated capacitive dilatometer of the strain cell, and $d_0$ was calculated from the measured data either by the value of strain at which the asymmetry parameter $\eta$ of the EFG tensor was zero, or by the minimum in the fits to the spin-lattice relaxation rate as a function of strain. The shear modulus of the epoxy was calculated from Young's modulus as measured by Hashimoto and Ikushima~\cite{Hashimoto_1980_MechanicalpropertiesStycast} as a function of temperature and assuming a Poisson ratio of $\nu = 0.36$.

\section{Notable BaFe$_2$As$_2$ Results}
\label{BaFe2As2_results}

\subsection{${}^{75}$As Spectral Measurements in the Antiferromagnetic State }
\label{sec:75As_spin-lattice_relaxation}

One major goal of this project was to use the advanced uniaxial pressure cells to determine if the predicted spin reorientation~\cite{Kissikov_2018_Uniaxialstraincontrol} occurs for high strain ($\varepsilon > 0.004$) in the magnetic state of BaFe$_2$As$_2$. Observation of such changes sets experimentially-verifiable checkpoints that can be used to determine a self-consistent theoretical understanding of the correlated electron behavior in the Fe-based superconductors. Additionally, understanding and control of spin polarization in a material could have practical applications in general in, for example, spintronics devices.

\begin{figure}[h!] 
\centering
    \includegraphics[trim=0cm 0cm 0cm 0cm, clip=true, width=0.5\linewidth]{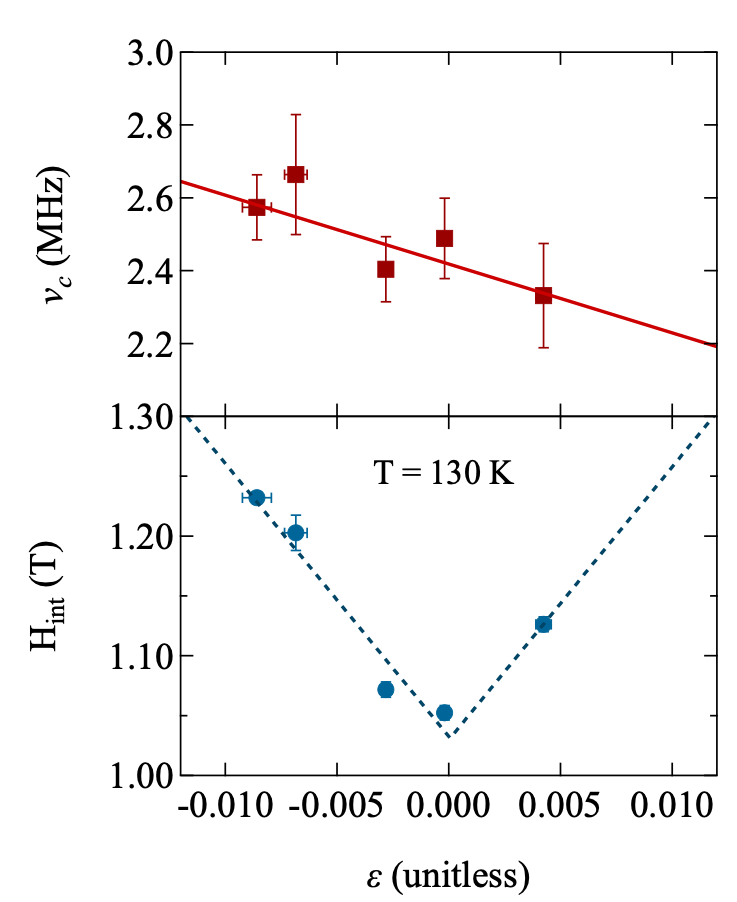}
    {\caption{\label{fig:n2_130K_Hint_and_nuC_vs_calib_strain_layout}(a) $c$-direction component of the EFG $\nu_c$ vs strain $\varepsilon$ at T = 130 K. (b) internal hyperfine field $H_\mathrm{int}$ vs strain $\varepsilon$. Solid and dashed lines are guides to the eye.}}
\end{figure}

In our preliminary data, we had observed no increase in internal hyperfine field at the ${}^{75}$As site for $T = 120$\,K, but based on further studies, it is more likely that the crystal had fractured for that run. These types of cell failures strongly motivated development of the unified control software pyStrainIFW mentioned above in Section\ref{sec:experimental_advancements}. Use of this software in later experiments allowed us to have a full history of all pressure cell parameters, allowing for easy diagnosis of numerous failure modes.

The preliminary $T = 130$\,K data set shown in the proposal was improved upon after calibration of the pressure cell, and we determined that up to a much higher maximum (corrected) strain of $\varepsilon = -0.009$, no spin-reorientation is evidenced by our NMR measurements. On the contrary, it seems that applied strain stabilizes the nematic/magnetic order up to higher temperatures and therefore  increases the constant-temperature value of the internal field as shown in Fig.~\ref{fig:n2_130K_Hint_and_nuC_vs_calib_strain_layout}. Additionally, we observe that the change in the electric field gradient can also be resolved in the nematic/magnetic state, as shown in Figure~\ref{fig:n2_130K_Hint_and_nuC_vs_calib_strain_layout}. 

Within the measured range, strain couples linearly to the hyperfine field, and therefore the magnetic moment on the Fe site in the nematic-orthorhombic/magnetic state. The monotonic increase of the internal field with strain along the orthorhombic $a$- or $b$-direction is consistent with the increase of the magnetic transition temperature with applied uniaxial pressure as measured via neutron diffraction~\cite{Tam_2017_Uniaxialpressureeffect}. This rules out a spin reorientation up to more than double previous maximum strain-dependent measurements  
\cite{Tam_2017_Uniaxialpressureeffect, Kissikov_2017_Localnematicsusceptibility, Kissikov_2018_Uniaxialstraincontrol}.

\subsection{Paramagnetic State ${}^{75}$As Spectral Measurements}
\label{sec:75As_spectral_measurements}

\begin{figure}[h!] 
    \centering
    \includegraphics[trim=0cm 0cm 0cm 0cm, clip=true, width=0.9\linewidth]{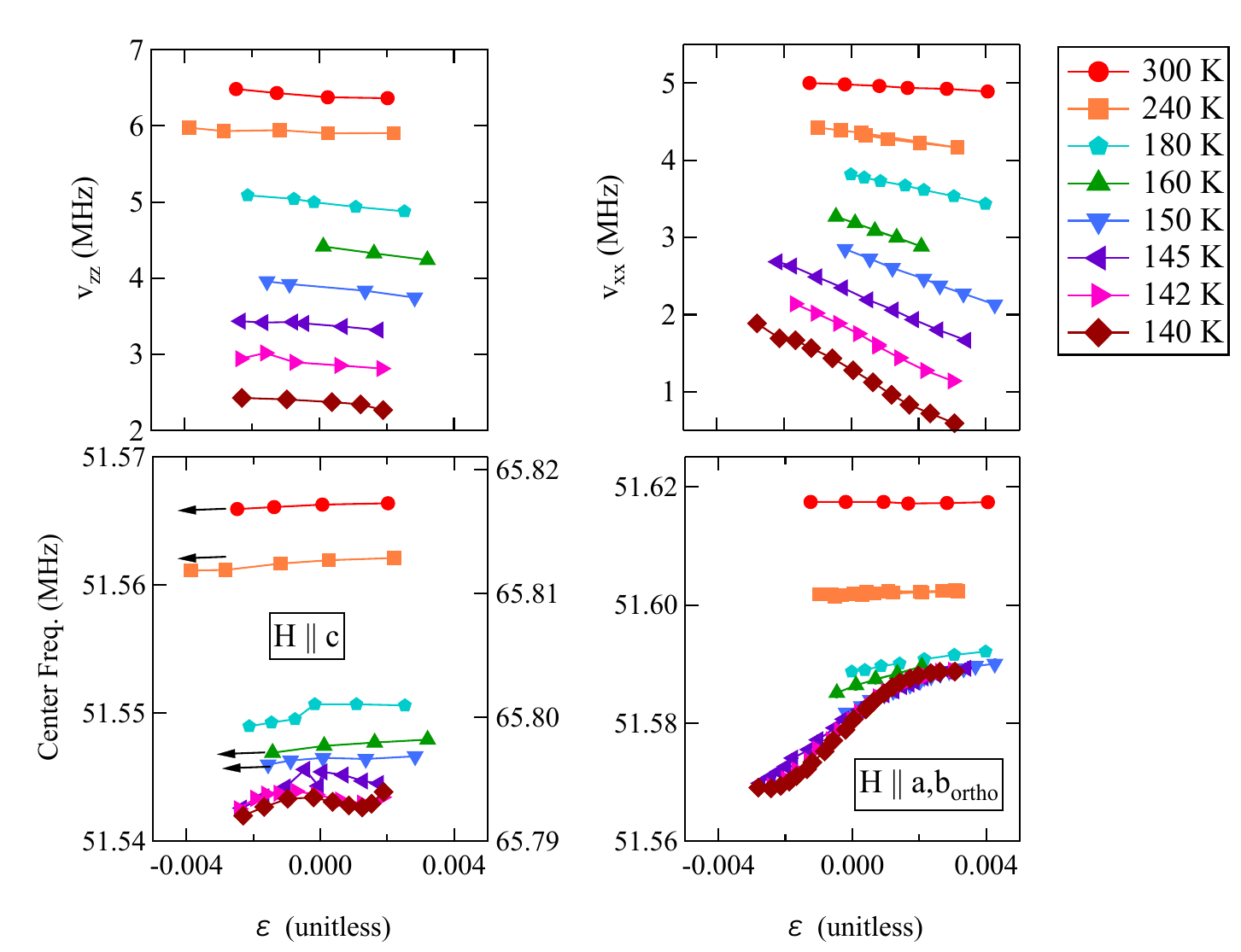}
    {\caption{\label{fig:s4p2_Vzz_Vxx_centfreqs_vs_strain}Extracted strain dependence of the electric field gradient (EFG) tensor components $\nu_{zz}$ and $\nu_{xx}$ for various temperatures from the same single crystal. The increasing magnitude of the slope with decreasing temperature of $\nu_{xx}$ is a measure of the nematic susceptibility.}}
\end{figure}

In addition to magnetic state measurements, we also performed a plethora of experiments in the
paramagnetic state of BaFe$_2$As$_2$. These determined that previous (unpublished) work did not
properly account for losses via the epoxy by which the sample is coupled to the strain cell. Our
experiments have served to both verify the published data \cite{Kissikov_2017_Localnematicsusceptibility} and identify further strain-dependent NMR parameters
 of relevance. In the paramagnetic state our spectral measurements as a function of temperature and
  strain elucidate two critical findings. First, by rotating the uniaxial pressure cell to have both
   $H \parallel \hat{c}$ and $H \parallel \hat{a},\hat{b}$, we were able to fully define the electric
    field gradient in a single crystal, whereas previous results required the assumption that the
     applied uniaxial pressure induced the same strain in two different crystals. Plots of the
      electric field gradient tensor principal components are shown in Figure~\ref{fig:s4p2_Vzz_Vxx_centfreqs_vs_strain}.

\begin{figure}[h!] 
    \centering
    \includegraphics[trim=0cm 0cm 0cm 0cm, clip=true, width=0.5\linewidth]{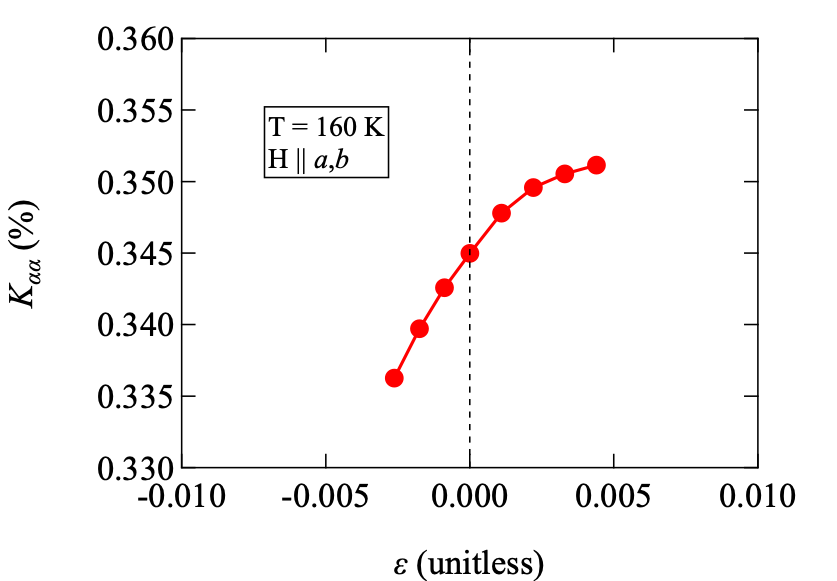}
    {\caption{\label{fig:BaFe2As2_K_vs_strain} Dependence of the NMR shift as a function of strain.}}
\end{figure}

Additionally, our precise spectral measurements allowed us to cleary show that Knight shift $K$ at the ${}^{75}$As site is strain dependent as shown in Figure~\ref{fig:BaFe2As2_K_vs_strain}. The origin of the non-linear dependence of the Knight shift is not yet understood, but may be a critical observation. The Knight shift is a measure of the local magnetic susceptibility at the ${}^{75}$As site, and therefore is sensitive to microscopic effects of the applied symmetry-breaking field. However, why the strain field would not couple linearly to the microscopic magnetic susceptibility is note clear, and this unexpected result motivates future theoretical investigation.

\subsubsection{BaFe2As2: Dynamics}
\label{sec:BaFe2As2_T1_T2}

The spin lattice relaxation rate $T_1^{-1}$, is a versatile tool to probe the low energy spin fluctuations in correlated electron materials. In the iron-based superconductors, the emergence of a electronic nematic state is preceded by strong fluctuations. As mentioned above, the lattice softens dramatically in the symmetry-breaking [110] and the elastic constant $C_{66}$\cite{Boehmer_2014_NematicSusceptibilityHole}, elastoresistance, NMR $T_1^{-1}$, and the EFG at the 75As site \cite{Kissikov_2018_Uniaxialstraincontrol, Kissikov_2017_Localnematicsusceptibility} were all shown to be a good probes of the diverging nematic susceptibility.

\begin{figure}[h!] 
    \includegraphics[trim=0cm 0cm 0cm 0cm, clip=true, width=0.5\linewidth]{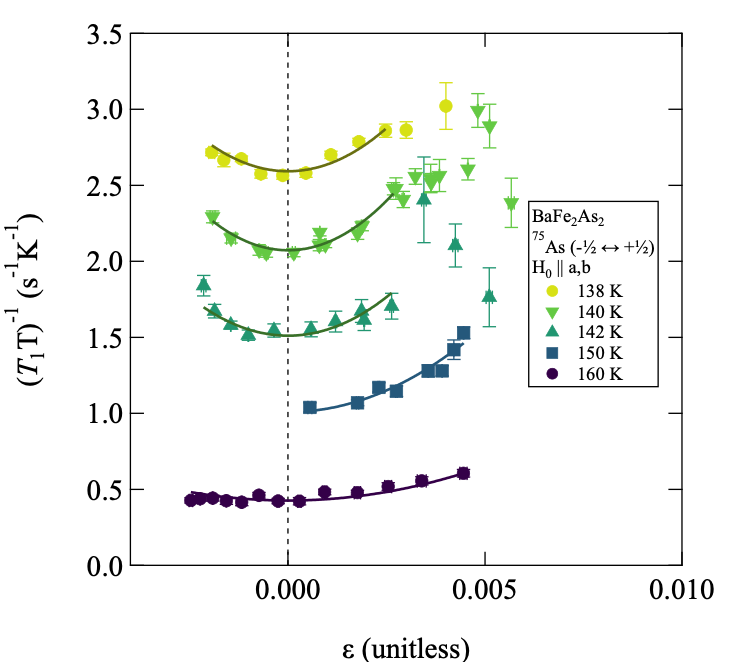}\includegraphics[trim=0cm 0cm 0cm 0cm, clip=true, width=0.5\linewidth]{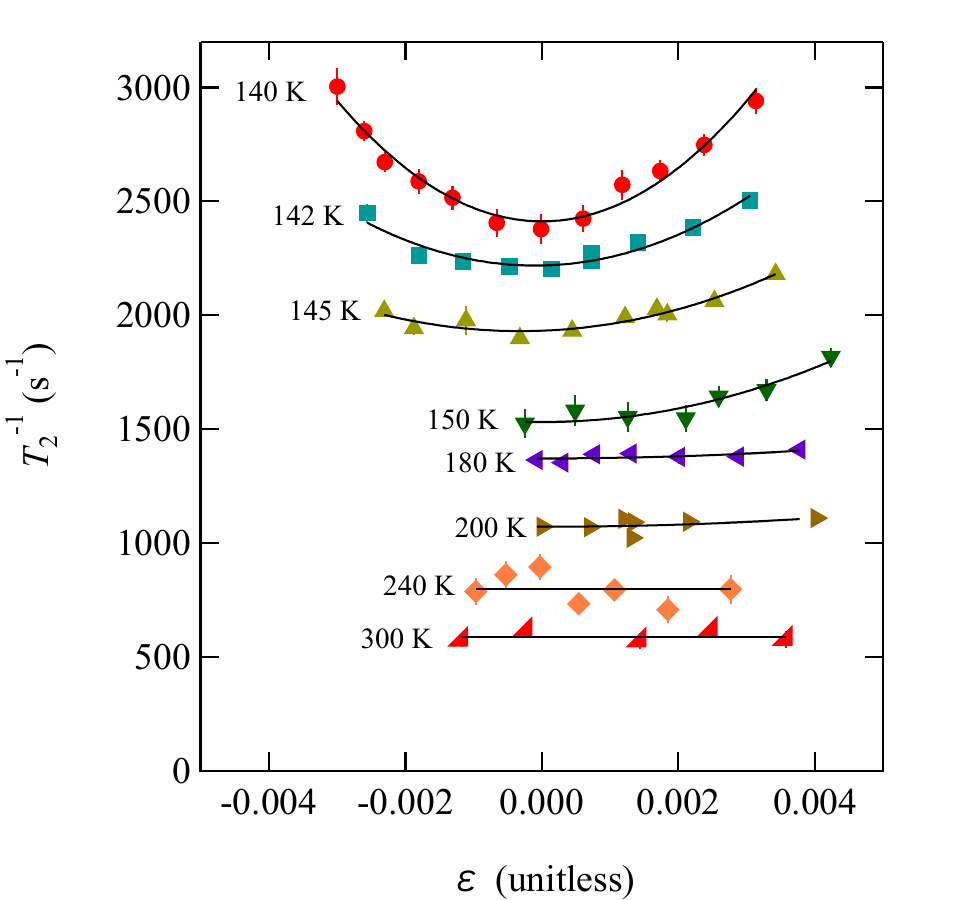}
    {\caption{\label{fig:BaFe2As2_T1_T2}(a) Strain dependence of the spin--lattice relaxation rate divided by temperature $(T_1T)^{-1}$ measured at the central transition of ${}^{75}$As at several temperatures above $T_N$ for external field applied in the basal plane at 45 degrees with respect to the tetragonal $a$-direction. (b) Strain dependence of the spin--spin relaxation rate $T_2^{-1}$. Both quantities couple to spin/nematic fluctuations and show increasing strain dependence with decreasing temperature above the nematic/magnetic transitions.}}
\end{figure}

Here we have expanded upon the existing work by also measuring the effect of strain on the spin--spin relaxation rate $T_2^{-1}$. Systematic studies of $T_2^{-1}$ were conducted as a function of temperature and strain show that $T_2^{-1}$, measured at the intense central transition is a also a good probe of the nematic susceptibility. Considering that T2 requires a fraction of the time to measure in comparison to T1, this allows for more in-depth studies in generalized electronic nematic materials.

\section{Closing}

As a direct result of this project, we have vastly improved the experimental NMR capabilities of the IFW, produced four high-quality and impactful peer-reviewed publications, and we mentored several masters, PhD, and bachelor students. In spite of the experimental challenges intrinsic to basic research, and a global pandemic, the funding from the DFG has made a significant impact. As discussed above, the results directly relevant to the proposed project are not insubstantial  We answered several of the proposed questions and a further publication is in preparation. 

I would like to acknowledge all those who helped with this research, particularly Dr. Hans-Joachim Grafe, but also Prof. Dr. Bernd B\"uchner, Dr. Piotr Lepucki,  Dr. Clifford Hicks, Dr. Saicharan Aswartham, and Dr. Sabine Wurmehl.

\section{Signature}

\begin{center}
\vspace{1cm}



............................................\\
\vspace{0.3cm}
Dr. Adam P. Dioguardi\\
Los Alamos, \today

\end{center}


\bibliographystyle{ieeetr}
\bibliography{DFG_final_report_Fe-based_strain}

\end{document}